%Paper: hep-th/9211011
%From: trivedi@theory3.caltech.edu (Sandip Trivedi)
%Date: Mon, 2 Nov 92 23:23:00 PST
%Date (revised): Mon, 9 Nov 92 15:04:03 PST

%\input apple
\tolerance=5000

\input phyzzx
\nopubblock
\titlepage
%\line{\hfill Preliminary Version}
\line{\hfill CALT-68-1833}
\line{\hfill DOE RESEARCH AND}
\line{\hfill DEVELOPMENT REPORT}
\title{SEMICLASSICAL EXTREMAL BLACK HOLES}
\author{SANDIP P. TRIVEDI\foot{Research supported in part by the U.S.
Department of
     Energy under contract no. DEAC-03-81ER4050. } }
\vskip.2cm
\centerline{\it Lauritsen Laboratory of High Energy Physics}
\centerline{\it California Institute of Technology}
\centerline{\it Pasadena, CA. 91125 }

\abstract{Extremal black holes are studied in a two dimensional model motivated
 by a  dimensional reduction from four dimensions. Their quantum corrected
geometry is calculated
semiclassically and a mild singularity is shown to appear at the horizon.
 Extensions of the geometry past the horizon are not unique
but there are continuations free from malevolent singularities.
A few comments are made about the relevance of these results to
 four dimensions and to the study of black hole entropy and information
loss. }
\endpage

\REF\davisa{P. C. W. Davis, Proc. Roy. Soc. London, A351(1976), 129. }
\REF\davisb{P. C. W. Davis, S. A. Fulling and W. G. Unruh, Phys. Rev.
  D13 (1976), 2720. }
\REF\hisa{W. A. Hiscock, Phys. Rev. D15 (1977) 3054. }
\REF\hisb{ W. A. Hiscock, Phys. Rev.
D16 (1977), 2673.}
\REF\birrela{N. D. Birrell and P. C. W. Davis, Phys. Rev. D18 (1978), 4408.}
\REF\andya{S. B. Giddings and A. Strominger, "Dynamics of Extremal Black
Holes," UCSB-TH-92-01. }
\REF\andyb{  B. Birnir, S. B. Giddings, J. A. Harvey and
   A. Strominger, Phys. Rev. D46(1992) 638.}
\REF\stana{J. G. Russo, L. Susskind and L. Thorlacius, "Black Hole
Evaporation in 1+1 dimensions," SU-ITP-92-4.}
\REF\stanb{ L. Susskind and L. Thorlacius,
Nucl. Phys. B382 (1992) 123.}
\REF\havrva{J. A. Harvey and A. Strominger, "Quantum Aspects of Black Holes,"
    Enrico Fermi Institute Preprint (1992). }
\REF\stevea{ S. B. Giddings, "Toy Models for Black Hole Evaporation,"
  UCSBTH-92-36. }
\REF\chiaraa{M. McGuigan, C. R. Nappi and S. A. Yost, "Charged Black Holes
In Two Dimensional String Theory," IASSNS-HEP-91/57. }
\REF\chiarab{ O. Lechtenfeld and
C. Nappi, "Dilaton Gravity and no Hair Theorem in Two Dimensions," IASSNS-
HEP-92-22.  }
\REF\lowe{D. A. Lowe, "Semiclassical Approach to Black Hole Evaporation,"
  PUPT-1340.}
\REF\banksa{T. Banks, A. Dabholkar, M. R. Douglas and M. O'Laughlin,
 Phys. Rev. D45 (1992) 3607.}
\REF\banksb{T. Banks and M. O'Laughlin, "Classical
and Quantum Production of Cornucopions at Energies Below $10^{18}$ Gev,"
 RU-92-14 (1992). }
\REF\hawkc{S. W. Hawking, Phys. Rev. Lett. 69 (1992) 406. }
\leftline {\bf Introduction:}

Hawking's discovery of blackhole radiance\REFS\hawka{S. W. Hawking,
Comm. Math. Phys. 43 (1975) 199. }\REFSCON\hawkb{S. W. Hawking, Phys.
Rev. D14 (1976) 2460. }\refsend raises several
intriguing questions. It shows that black holes have an entropy which
can be elegantly expressed in terms of their geometry but which
remains mysterious in terms of any underlying
microstates. It also suggests that because of the thermal nature of the
outgoing radiation,
a  loss of  quantum coherence might occur in processes involving black holes.
 Extremal black holes provide a convenient
setting in which to address both these issues.
Their zero temperature suggests that their entropy should be explained
in terms of a degeneracy of ground states. It also
 makes them convenient toy laboratories in which to study scattering and
 a potential loss of quantum coherence.

  In this paper, we study a model consisting of dimensionally reduced
gravity and electromagnetism coupled
to two dimensional
scalar fields. Classically, this model has Reissner Nordstrom black
hole solutions, obtained from dimensionally reducing the usual four dimensional
charged black hole solutions. In this paper, we concentrate for the most part
on the extremal black holes which in Planck units have a mass equal
to their charge and have zero Hawking temperature.  We show  that contrary
to expectations, the
vacuum polarisation of a scalar field in the background of such an
extremal black hole blows up at the horizon.
This raises the possibility of their geometry being drastically
modified in the vicinity of the
horizon and their entropy
being very different from what classical considerations would suggest.
A semiclassical calculation of their quantum corrected geometry shows
however, that this is not true.
  For large
black holes the value of the dilaton at the horizon and hence their
entropy \foot{ The entropy of these black holes
depends on the dilaton and is large if the value of the dilaton at the
horizon is large.}, stays large. A singularity does appear at
the horizon but it
is very mild. For example,
 tidal forces and the  curvature stay finite at the  horizon.
 This suggests that there should be a continuation of the geometry
 past the horizon. In fact, there is more than one such continuation -
even when we restrict ourselves to static solutions.
In one of these the causal structure of
space time  is much like the classical extremal
solution. But there  is another continuation possible
 in  which
the causal structure is much different and in which  there are no
malevolent singularities. We conclude with a brief discussion of the
relevance of our results to the study of four dimensional extremal
black holes, and to the study of black hole entropy and information loss.

The study of quantum effects in  two dimensional black holes was
first  undertaken in the
$ 1970 's $ in some very interesting papers which include
references [1-5]. More recently, considerable interest was renewed by the
discovery of  a two dimensional black hole solution in
the work of Mandal, Sengupta and Wadia \Ref\tata{G. Mandal, A. Sengupta
and S. Wadia, Mod. Phys. Lett. A6(1991) 1685.}
and
Witten. \Ref\witten{E. Witten, Phys. Rev. D44(1991) 314.}
This solution was then used to study questions related to Hawking
evaporation in the work of Callan, Giddings, Harvey and Strominger
\Ref\cghs{C. G. Callan, S. B. Giddings,
J. A. Harvey and A. Strominger, Phys. Rev. D45 (1992) 1005.} (CGHS).
Two recent papers which review the subsequent developments are references
[10,11].
Papers especially relevant to the work presented here
include references [6,7,8,9,15,17] .  Papers which discuss dimensionally
reduced models
include references [12,13,14] . To our knowledge, the first published reference
to the idea of using extremal black holes for studying issues related
to Hawking
radiation is in the paper of Preskill et. al.\Ref\extremal{J. Preskill,
P. Schwarz, A. Shapere, S. Trivedi and F. Wilczek, Mod. Phys. Lett. A6 (1991)
2353. }.

\leftline {\bf The Model:}

We start  in four dimensions
and make a spherical symmetric ansatz for the metric which gives
$$ ds^2=  g_{\alpha \beta} dx^{\alpha} dx^{\beta} + e^{-2 \phi} d \Omega .
                                   \eqn\sphericalmetric  $$
Here $ g_{\alpha \beta} $ is the two dimensional metric in the
"r-t" plane and $e^{-2 \phi}$, which we call the dilaton, is the  square of
the radius of the two sphere.
The Einstein Hilbert action then takes the form
$$ S_G= {1 \over 4 G }
          \int d^2x \sqrt{g} e^{-2 \phi} \left( R + 2 (\nabla \phi)^2
              + 2 e^{2 \phi} \right ) .       \eqn\gravityaction  $$
We note that this differs from the action considered by CGHS \refmark{\cghs }
 in the
form of the dilaton potential, i.e. the last term above. The term considered
here prevents the action $S_G$ from scaling simply under the transformation
$ \phi \rightarrow \phi + c $.

Proceeding similarly, the Maxwell field can be dimensionally
reduced to give  an action
$$S_{EM} = - { 1 \over 4 G }
              \int d^2x \sqrt{g} e^{-2 \phi} F^2 \eqn\maxwellaction $$
where $F^2 $ now refers to the field strength of a two dimensional gauge field.

  Spherically symmetric charged black hole solutions of the original
four dimensional theory continue to be solutions of this theory.
They are given by a metric
$$ ds^2 = -(1- {2M \over r} + {Q^2 \over r^2 }) dt^2
                + {1 \over (1-{2M \over r} + {Q^2 \over r^2})}dr^2 \eqno\eq $$
and a dilaton
$$ e^{-2 \phi} = r^2 .  \eqno\eq $$

The corresponding field strength is

$$ F_{r t} = {Q \over r^2 }.       \eqno\eq $$

Here, $M$ is the mass and $Q$ the charge of the black hole.
We will  be mainly interested here in  extremal black holes for which
 $M=Q$.

     In order to incorporate quantum effects in a manageable way we
couple the above theory to N scalar fields.
%In the interest of tractability,
% we take the scalar fields to be  conformally free scalar
%fields in two dimensions. Unfortunately, as a result this model
%does not retain an
%obvious four dimensional interpretation.
The parameter N allows us to
consider the theory in the large N limit in which $ N \rightarrow \infty $
and $\hbar \rightarrow 0$ while  keeping $ N \hbar $ fixed. This allows us to
systematically incorporate the quantum effects due to scalar loops
-  which go as
$N \hbar $ - while keeping the other fields classical. The scalar fields are
taken to be conformally coupled scalar fields in two dimensions. This is done
in the interest of tractability and the model we consider here, with
electrically charged black holes, does not retain an obvious four dimensional
interpretation. However, all our conclusions will go through, unchanged,
for a closely related model obtained by dimensionally reducing a system
consisting of fermions coupled to a magnetically charged black hole in four
dimensions \foot{I would like to thank
A. Strominger for pointing this out
to me.}. The scalar fields will then correspond to the bosonised version
of the Callan-Rubakov modes of the fermions\refmark{ \andya, \banksa }.
We should add though, that
from a strictly four dimensional point of view, even in this model,
the other modes
of the fermion fields will contribute to the quantum stress
tensor and their neglect cannot be justified.

\leftline {\bf  Vacuum Polarisation:}
We intuitively expect quantum effects associated with curved spacetime
to be small if the curvature is small. One way to make this intuition
more precise is to integrate out the scalar fields and look at the induced
action along with the original action of Einstein gravity. In conformal
gauge this would be given  by

$$ \eqalign{S_{IND} + S_G  =  {1 \over 4 G } \Bigl [  \int d^2x
                            \bigl \lbrace   e^{2 \rho }
                & - 4  e^{- 2 \phi}    \partial_{+}(\phi + \rho)
                                        \  \partial_{-}(\phi + \rho) \cr
      &   + 4 e^{-2 \phi} (1 -\ G {N \hbar \over 12 \pi} e^{2 \phi} )
                   \   (\partial_{+} \rho
                       \partial_{-} \rho ) \ \bigr \rbrace
                    \ \Bigr ] \cr}           \eqno\eq   $$

We  see that there is  critical value of the dilaton
field given by
$$ e^{-2 \phi} = G {N \hbar \over 12 \pi } \eqn\critical $$
at which the Liouville kinetic energy term becomes degenerate.
But if $ e^{- 2 \phi} $ is
much larger than this critical value the induced term
should have a small effect.
%$$ {N \over 24} \hbar \ G \ e^{2 \phi } \ll 1 \eqno\eq $$
For a black hole with a mass much bigger than the Planck mass this is true
from asymptotic infinity all the
way  up to the horizon. Thus we would expect quantum effects to be small in
this region. In fact, as we show below, the vacuum polarisation for a massless
scalar field diverges at the horizon of an extremal black hole
no matter how large it's mass.
%One legitimate concern in this regard is to ensure that  we calculate the
%stress tensor in a physical quantum state of the scalar field. It is
%quite easy to show that for an extremal black hole the Boulware
%Unruh and Hartle - Hawking states are all the same ( we will
%refer to this as the BUHH state ). But to remove all doubt
%we carry out a general analysis without restricting ourselves to any
%particular state and thereby show that even the most well behaved stress
%%tensor
% is divergent.

We work in Schwarzschild gauge where the metric
is given by
$$ ds^2 = -f dt^2 + {1 \over f} dr^2  \eqn\schwarzgauge $$
The conservation equations for the stress tensor then imply that \Ref\candf{
S. M. Christenson and S. A. Fulling, Phys. Rev. D15(1977) 2088. }
$$ T^r_t = c_1 \eqn\stressrt $$
and that
$$ T^r_r={1 \over  2 f } \int_{r_h}^r f^{'} T^{\mu}_{\mu} dr + {c_2 \over f }
                  ,  \eqn\stressrr $$
where $r_h $ refers to the position of the horizon.
An ambiguity in state of the scalar field is
reflected in the two arbitrary constants $c_1$ and $c_2$. Consider now an
 observer freely falling into the black hole with a four velocity
$({p_0 \over f}, -\sqrt {p_0^2 -f})$.
 She sees an energy density
$$T_{\mu \nu} U^{\mu} U^{\nu} = T^r_r ({p_0^2 \over f } -1) -
                                 T^t_t {p_0^2 \over f }
                                  -2 T^r_t {p_0 \over f^2 }
                 \  \sqrt {p_0^2 -f }. \eqno\eq $$

Substituting from equations \stressrt \ and \stressrr \  we see that
$$ \eqalign{T_{\mu \nu} U^{\mu} U^{\nu} = & {2 (c_2 -c_1) p_0^2 \over f^2} +
                                (-T^{\mu}_{\mu} + {1 \over f} \int^r_{r_h}
                              f^{'} T^{\mu}_{\mu} dr) { p_0^2 \over f} \cr
                             &   -{(c_2 -c_1) \over f} -
                           {1 \over 2f} \int^r_{r_h} f^{'} T^{\mu}_{\mu} dr
                                         . \cr}
                               \eqn\stressenergy $$
Where $f^{'}$ refers to the derivative of $f$ with respect to $r$.

This shows that the leading divergence goes as
${(c_2 -c_1) \over f^2 } p_0^2 $. So we restrict  ourselves to   states
 in which $c_2=c_1$.
Then, using the trace anomaly
$$ T^{\mu}_{\mu} = -{ f{''} \over 24 \pi }, \eqn\traceanomaly $$
and L 'Hospital's rule
 we see that close to the horizon
%$$ T^{\mu}_{\mu} = -f{''}/{ 24 \pi } \eqno\eq $$
%we have that the energy density above  goes like
%$$ T_{\mu \nu } U^{\mu} U^{\nu} = {1 \over 24 \pi}
%                        {({f^{'})^2 \over 2 }- f f^{''} ) \over f^2}
%               \  p_0^2
%         -{1 \over 4 f } (f^{'})^2 . \eqno\eq $$
%Thus close to the horizon the
$$ T_{\mu \nu } U^{\mu} U^{\nu} \simeq
                      const \ { f^{'''} \over f^{'} }. \eqn\hordiv $$

And this diverges since for an extremal black hole $f$ has
a double zero at the horizon.
In other words, if $ \delta \tau $ is the proper time taken to reach the
horizon the
energy density diverges like

$$ T^{\mu}_{\nu} U^{\mu} U^{\nu} \sim 1/( \delta \tau ). \eqn\div$$

  We note that the  analysis above was very general without restrictions to
any particular state of the scalar field. In fact for these black holes,
 the Boulware, Unruh and Hartle-Hawking states are the same
and considerations similar to those above at the past horizon,
would single it out as having the minimal divergence.

  This divergence can be better understood by regarding an extremal black hole
as the limit of a non extremal one. A non extremal black hole has an outer
and an inner horizon and these come together in the extremal limit. Let us
take $r_h$ in equation \stressrr \ to refer to the outer horizon. Then
as before, we see that $c_1$ must equal $c_2$ for the leading divergence
to  vanish at the outer horizon. Furthermore, equation \hordiv \ shows that
the stress tensor stays finite at the outer horizon, since $f$ has
a single zero. Now let us focus on the inner horizon. Since $r_h$ in
equation \stressrr \ was taken to mean the outer horizon and $c_1=c_2$ we see
from equation \stressenergy \ that the leading divergence goes as
$$ T_{\mu \nu} U^{\mu} U^{\nu} \simeq - {p_0^2 \over 48 \pi}  \
                {\bigl [f^{'}(r_{inner})^2 - f^{'}(r_{outer})^2 \bigr ]
                            \over f^2 }.
                  \eqn\divnon $$
And this diverges since the quantity within brackets does not cancell and
$f$ is zero - albeit a simple zero- at the inner horizon. If $ \delta \tau $
is the proper time taken by a freely falling observer to reach the horizon
this implies that
$$ T_{\mu \nu} U^{\mu} U^{\nu} \simeq {const \over (\delta \tau)^2 }.
                      \eqn\divnonb $$

      In summary, we find that if we adjust the quantum state of the scalar
field so that the stress tensor is finite at the outer horizon, it diverges
at the inner horizon. It is perhaphs not so surprising then, that in the
extremal case when the two horizons come together the divergence
persists, although in a softened form.

\leftline{\bf Quantum Corrected geometry :}
We work in "Schwarzschild" gauge equation \schwarzgauge \ and look for
static  solutions  which incorporate  the vacuum polarisation.
The equation of motion of the Maxwell field gives
$$ F^{r t} = {Q \over e^{-2 \phi} } . \eqno\eq $$
The equation obtained by varying the trace of the metric then becomes :
$$ \bigl [f (e^{-2 \phi})^{'} \bigr ]^{'} - 2 + 2 {Q^2 \over (e^{-2 \phi})}
= -\xi \ f{''} . \eqn\metricb $$
Here
$$ \xi = G \ { N \hbar \over 48 \pi }. \eqn\metricc $$

Similarly the equation obtained by varying the dilaton is
$$ -f^{''} = {1 \over 2} f {[(e^{-2 \phi})^{'}]^2 \over (e^{-2 \phi})^2}
             + \bigl [ {f (e^{-2 \phi})^{'} \over e^{- 2\phi} } \bigr ]^{'}
                -{2 Q^2 \over (e^{-2 \phi})^2 }. \eqn\dilaton $$
Finally, the $T_{++}$ and $T_{--}$ constraints tell us that

$$ -{1\over 4 } f^2  \bigl [ (e^{-2 \phi})^{''} -{1 \over 2 }
                     { (e^{-2 \phi}){'}^2 \over e^{-2 \phi} } \bigr ]=
                           {\xi \over 4}[ f f^{''} - {1\over 2} (f^{'})^2 ]
                                        +C. \eqn\metrica $$
Here, C is an arbitrary constant which indicates an ambiguity in the
quantum state of the scalar fields. We expect extremal black holes
to have zero temperature and seek
solutions with $C = 0 $ and $ f^{'} =0 $ at the horizon ( where $ f=0 $ ).
If $x$ represents the coordinate distance from the horizon  this
suggests that close to the horizon:
$$ f \simeq \alpha_1 x^2 + \alpha_2 x^{3 + \delta} \eqn\formf $$
and
$$ e^{-2 \phi} \simeq d_h + d_2 x^{ 1 + \delta}. \eqno\eq $$
Equations \metricb \ and \dilaton \ then show that
$$ Q^2 ={ d_h^2 \over \xi + d_h } \eqno\eq $$
and that
$$ \alpha_1 = {1 \over \xi + d_h}. \eqno\eq $$
Further equation \metricb \ then shows that
$$ d_2 = - \xi {\alpha_2 \over \alpha_1 }
                   {( \delta + 2 ) \over \delta }. \eqno\eq $$
  Finally equation \dilaton \  shows that delta is given by the equation
$$ \delta = {\bigl (-3 + \sqrt{9 + 24 {\xi \over ( d_h -\xi) } } \ \bigr )
                            \over 2}.
                      \eqn\findel $$
These values of the parameters can also be shown to be consistent with
equation \metrica \ (with $C$ set equal to $0$).
Like their classical counterparts, these solutions have
only one free parameter, which we can take to be   $d_h$,  the
value of the dilaton at the horizon. $ \vert \alpha_2 \vert $ can be set
equal to $ 1 $ by  rescaling $x$. The other parameters are then
uniquely determined. Numerical computations show that with $ \alpha_2  =
-1 $ the solution evolves to an asymptotically flat geometry
as $ x \to \infty $. As a special case of equation \findel \ note
that for large black holes  where $d_h \gg \xi $, $ \delta \simeq 2 \xi/d_h $ .

\leftline{\bf Discussion:}

   What do we learn from these solutions? Extremal black holes are
 dangerously close to becaming naked singularities. And
we might have
thought that the diverging stress tensor would cause a singularity to appear
at the horizon and even drive the dilaton to it's critical value
equation \critical \ ,
thereby changing the entropy of the black hole dramatically. However, this
does not happen. The value of the dilaton field at the horizon remains a
free parameter and for large blackholes ($ d_h \gg \xi $ ) the entropy stays
large and remains as mysterious as ever.

For generic values of $ d_h$, $ \delta$ is not an integer and the solution is
non analytic in $x $.
This non analyticity implies a very mild
singularity at the horizon. The second derivative of the curvature
as seen by a freely falling observer
diverges as she crosses the horizon, but the tidal forces and the curvature
stay finite. Thus there should be an extension of the geometry
 past the horizon. In fact there  are two obvious static extensions.
These correspond to replacing
$ x^{3 + \delta } $  in equation \formf \ above by
         $ \vert x \vert ^{3 + \delta}  $
or by $ x^3 \vert x \vert ^{\delta}  $ and correspondingly extending the
dilaton field. We will call these the even and odd extensions respectively.
It can be shown that both of these satisfy the junction conditions which
arise from equations \metrica \ , \metricb \  and \dilaton \ .
\foot{
For special values of $ d_h $ ( within the semiclassical regime ),
$ \delta $ becomes an integer and  depending on it's value either of
these can  became the
analytic continuation, in which
 even the mild singularity disappears.}
The odd continuation results in a spacetime much like the classical extreme
black hole spacetime in which we can hit a time like singularity within
a finite proper time of crossing the horizon. \REF\horowitz{J. H. Horne
and G. T. Horowitz, Nucl. Phys. B368(1992) 444. }
The even continuation
though, results in an entierly different spacetime, in which we
pass from one asymptotically flat universe to another without
encountering any  malevolent singularities at all!\foot{A spacetime with the
same causal properties has been found by Horne and Horowitz.\refend }
The corresponding
Penrose diagrams are shown in figures 1 and 2 respectively.

We have not investigated as yet, whether any of these
extensions are relevant for a blackhole formed from collapse; but it is
not inconceivable that at least some part of spacetime outside
 a collapsing "star" is
described by either of these. In this case it should be
possible to begin the collapse from one asymptotically flat universe
and open out into another.
Information thrown in from one asymptotically flat universe could then
end up in another. Indeed such black holes would be the "ultimate" remnants.
\refmark{\cghs, \banksa , \banksb}
%\Ref\banks{T. Banks, A. Dabholkar, M. R. Douglas and M. O'Loughlin,
%Phys. rev. D45 (1992) 3607;T.Banks and M. O'Laughlin, "Classical and
%Quantum Production of Cornucopions at energies below $10^{18}$ Gev",
%Rutgers preprint RU-92-14 (1992).}.
Information would not just be hiding in some long tube but would have found
it's way into another universe and be lost forever. This model might also
furnish a simple context in which to explore the production of such remnants
inexternal fields .

\REF\His{ P. Anderson and W. Hiscock, Private communication }
We cannot say much  about four dimensional black holes,
since
   we do not know how the stress tensor of a four dimensional scalar
field behaves\foot{ These calculations are currently in progress \refend .}
.  If we do interpret the two dimensional
metric and the dilaton as components of a spherically symmetric  four
dimensional metric we find that the tidal forces and the curvature  at
the horizon stay finite. This  suggests that if the four dimensional
stress tensor  behaves similarly to the two dimensional case and
blows up for example, no faster than $ { 1 \over \delta \tau } $
(equation\div \ );
a very mild singularity would form at the horizon. In particular
the area of the horizon would stay large and so would the entropy.

  Is there a loss of information in scattering quanta off these
extremal black holes? Unfortunately, we cannot answer this question
conclusively  here.
It is clear that
some information regarding  the kinds of scalar quanta thrown
in will be lost in scattering. But, in the large $ N $ limit
considered here the entropy
of the black hole - which goes as $ 1/\hbar $ - is large\foot{ I would like to
thank J. Preskill for pointing this out and for the subsequent argument about
the large $N$ limit being inadequate.}. If this entropy has an explanation
in terms of underlying microstates, this would suggest a large degeneracy
of ground states. And we cannot exclude the possibility that the lost
information is hiding in correlations between the ground states and
the outgoing radiation. To settle this issue would
require keeping track of  very subtle
correlations ( beyond
leading order in $ N $ ) in the outgoing radiation.
We hope, in subsequent work to return to this problem.

Finally, a few comments regarding non extremal
black holes in contact with a heat bath.
%We have studied static solutions corresponding to black holes
%in contact with a heat bath.
 Classically these black holes have an outer and inner horizon.
Numerical calculations show that the behavior of the geometry
inside the outer horizon depends, for a given value of the dilaton field,
on the electric charge. If the electric charge is small the dilaton decreases
in value till it reaches it's critical value equation \critical \ and a
space like singularity forms.
 However once the charge is large enough the dilaton does not
go to it's critical value. Instead an inner horizon forms
 at which the  metric component
$ f$ ( equation \schwarzgauge \  )  behaves like
$$ f \simeq \alpha_1 \ x - {\alpha_1 \over 2 } { x \over log(x)} \eqno\eq $$
and the dilaton behaves like
$$ e^{-2 \phi} \simeq -const \  log(x)  \eqno\eq $$
where $x$ is the coordinate distance from the horizon.
\REF\ori{A. Ori, Phys. Rev. Lett. 67(1992) 789.}
This shows that the curvature  (which is related to the second derivative
of f)  and hence the tidal forces as felt by a
freely falling observer blow up at the inner horizon. But it is
straightforward to show that the divergence is mild enough for
the tidal impulse to stay
finite, which suggests
\foot{A. Ori has found a similar divergence in his study
of mass inflation in four dimensions. \refend }
 that there should be an extension of the geometry
past the inner horizon.
 If
the metric and dilaton  above are regarded as components of a
spherically symmetric
metric in four dimensions though,
there are components of the tidal impulse that blow up ( although rather
slowly), which  suggests that the four dimensional black holes
might behave differently.

\leftline{\bf Acknowledgement :}
It is a pleasure to acknowledge the help M. Bucher, G. Horowitz,
 J. March-Russell, E. Poisson,
A. Ridgway, K. Thorne, E. Verlinde,  E. Witten and P. Yi have given.
This work greatly benefitted from conversations with
J. Preskill, A. Strominger and
F. Wilczek and it is a special pleasure to thank them
for  their generous insights,
and encouragement.

\endpage
\refout
\bye